\documentclass[aps,preprint, 11pt,nofootinbib]{revtex4}
\usepackage{graphics}
\usepackage{slashed}
\usepackage{amssymb}
\usepackage{lscape}
\usepackage{amsmath}
\usepackage{rotating}
\usepackage{graphicx}
\graphicspath{ {images/} }
\begin{document}

\title{Quantum magnetic monopoles at the Planck era from unified spinor fields}

\author{$^{2,3}$ Mauricio Bellini
\footnote{{\bf Corresponding author}: mbellini@mdp.edu.ar}}
\address{$^2$ Departamento de F\'{\i}sica, Facultad de Ciencias Exactas y
Naturales, Universidad Nacional de Mar del Plata, Funes 3350, C.P. 7600, Mar del Plata, Argentina.\\
$^3$ Instituto de Investigaciones F\'{\i}sicas de Mar del Plata (IFIMAR), \\
Consejo Nacional de Investigaciones Cient\'ificas y T\'ecnicas
(CONICET), Mar del Plata, Argentina.}

\begin{abstract}
I use Unified Spinor Fields (USF), to discuss the creation of magnetic monopoles during preinflation, as excitations of the quantum vacuum coming from a condensate of massive charged vector bosons. For a primordial universe with total energy $M_p$, and for magnetic monopoles created with a total Planck magnetic charge $q_M=q_P=\pm e/\sqrt{\alpha}$ and a total mass $m_M$, it is obtained after quantisation of the action that the fine-structure constant is given by: $\alpha= \frac{5}{6}  \left(1- \frac{16 \,m_M}{5 \,M_p}\right) \,\left(\frac{e}{q_M}\right)^2$. If these magnetic monopoles were with total magnetic charge $q_M=\pm e$ and a small mass $m=m_M/n$, there would be a large number of small quantum magnetic monopoles which could be candidates to explain the presence of dark matter with a $30.97\,\%$ of the energy in the primordial universe at the Planck era. The case of milli-magnetically charged particles is also analysed. We demonstrate that magnetic monopoles (MM) with masses less than $3.6\times 10^3$ GeV, can exist with a very small charges of up to $10^{-14}\,e$, which are quantities of interest for searches to be performed in the ATLAS and MoEDAL experiments.
\end{abstract}

\maketitle

\section{Introduction and motivation}

Dirac's quantisation\cite{dirac,dirac1} is consistent with the existence of magnetic monopoles (DM), where the magnetic charge is related to the electric charge. The discovery for signatures of primordial monopoles, especially superheavy ones, would have consequences for particle physics and cosmology. There are topologically stable magnetic monopoles\cite{tof,poly}, which are related to quantum of Dirac magnetic charge. This kind of MM are predicted by all the standard grand
unified theories (GUT). There are theoretical arguments which suggest that these monopoles are not expected to be present in the galaxy. On the other hand, the implementation inflationary scenarios\cite{infl,infl1,infl2} in
GUTS leads to the conclusion that these monopoles\cite{la,oli,gut} are inflated away and the reheating temperature at the end of inflation is close to $10^{10}$ GeV, which is very low to produce superheavy monopoles during the thermalisation process. However, the possibility that these monopoles were created without thermalisation deserves be considered\cite{ruso}, because describes an alternative explanation for the MM creation from extra dimensions. Various searches have been carried out utilising diverse detection techniques in both observational facilities and experiments in colliders. Neutrino telescopes, such as ANTARES and IceCube currently, with KM3NeT and PINGU in the future, pursue the search for GUT monopoles in the cosmic front. The CERN LHC, is the ideal machine to produce MM, if they exist. In particular the MoEDAL experiment is specialising in the detection of MM and other highly ionising particles. MoEDAL has set the stringent bounds on high magnetic\cite{A} and together with LHC experiments, such as ATLAS\cite{B} and CMS, are going to continue probing the existence of light monopoles. Volumes exposed to LHC collisions, such as the Run-1 CMS beam pipe\cite{C}, will be scrutinised by SQUID for the presence of trapped monopoles. Another possibility to probe MM is given by their bound state; the monopolium, which can be tested in ATLAS and CMS. On the other hand, candidates for non-baryonic dark matter (DM) must satisfy several conditions: they must be stable on cosmological time scales and must interact very weakly with electromagnetic radiation. Furthermore, they must have the right relic density. Candidates include primordial BH, axions, sterile neutrinos, and weakly interacting massive particles (WIMPs). In particular WIMPs are particles with mass between $10\, {\rm GeV}$ and a few TeV, with cross sections of weak strength\cite{D}.

In this work we shall study the origin of MM as quantum excitations of the vacuum of very massive and charged spin 1-bosons, which are topologically stable because the existence of structure coefficients originated by the quantum algebra that comply these bosons. To study this issue we shall use the Unified Spinor Fields (USF) theory, which was introduced in some recent works\cite{usf1,usf2,usf3}. Because we are interested in the study of this phenomena at the beginning of the universe we shall consider the preinflationary scenario\cite{pr,pre,mb1}. This scenario describes the origin of the universe when it has not yet thermalised, using a global topological phase transition from a 4D Euclidean manifold to an asymptotic 4D hyperbolic one. The interesting of this idea is that $\tau$ is a space-like coordinate before the big bang, so that can be considered as a reversal variable. However, after the phase transition it changes its signature and then can be considered as a causal variable. The work is organized in the following manner: in Sect.II we describe the preinflationary model by making emphasis in the classical background dynamics. In Sect. III, se revise the boundary conditions, when it is varied the Einstein-Hilbert action, we describe the Ricci flow, we revise the covariant derivatives on the extended manifold, which describes an "elastic" geometry in which there is nonzero non-metricity. Furthermore, we introduce the quantum space-time and the line elements used in the theory (one for the space-time coordinated and the another for the inner space-time). In Sect. IV we introduce the action that describe the excitations of very massive and charged vectorial bosons the originates the MM, we calculate the expectation value for the action and we estimate the total mass of these monopoles at the beginning of the universe, and the possible existence of MM with small mass and a fractional charge of $e$. Also, we provides the dynamical equation that describes these boson fields during the preinflationary expansion of the universe. Finally, in Sect. V we develop some final remarks and conclusions.

\section{Preinflationary dynamics}

A theoretical possibility suggested recently is that during preinflation the universe begun to expand through a (global) topological phase transition\cite{pre}. In this model was proposed a model to describe the birth of the universe using a complex time: $\tau(t)=\int e^{i \theta(t)}dt$, where the transition from a preinflationary to an inflationary
expansion was described by using a dynamical rotation of the complex time, $\tau(t)$, on the complex plane, being the spacetime described by a complex manifold. Furthermore was defined the dynamical variable $\theta$:
$\pi/2 \geq \theta(t)>0$, which is related with the expansion of the universe
\begin{equation}
\theta(t)=\frac{\pi}{2} e^{-H_0 t}.
\end{equation}
The line element introduced in\cite{mb1} to describe preinflation, is
\begin{equation}\label{m}
d{x}^2 =  \left(\frac{\pi a_0}{2}\right)^2 \frac{1}{\theta^2} \left[{d\theta}^2 - \delta_{ij} d{x}^i d{x}^j\right],
\end{equation}
where $\theta_0=\frac{\pi}{2}$. During the phase transition, the universe initially is described by a Euclidean metric, but later evolves to a globally asymptotic hyperbolic space time: $\theta {\rightarrow } 0$. The Lagrangian density for the scalar field:
\begin{displaymath}\label{lan}
{{\cal L}}_{\varphi}= -\left[\frac{1}{2} g^{\alpha\beta} \varphi_{,\alpha}\varphi_{,\beta} - V(\varphi)\right].
\end{displaymath}
The stress tensor related to ${{\cal L}}_{\varphi}$, is
\begin{equation}
{T}_{\mu \nu}=-\left[\varphi_{,\mu}\varphi_{,\nu}-g_{\mu \nu}\left(\frac{1}{2} g^{\alpha\beta}\varphi_{,\alpha}\varphi_{,\beta}-V(\varphi)\right)\right],
\end{equation}
We are dealing with a spatially isotropic and homogeneous background metric (\ref{m}), and therefore the scalar field only depends on time, and complies with the dynamics
\begin{equation}\label{dy}
\varphi'' + 2\,{\cal H}(\theta)\varphi' + \frac{\delta V(\varphi)}{\delta\varphi}=0,
\end{equation}
where the prime denotes the derivative with respect to $\theta$ and ${\cal H}=a'/a=-1/\theta$ is the conformal Hubble parameter, for a conformal scale factor
\begin{equation}
a(\theta)=\left(\frac{\pi a_0}{2}\right) \frac{1}{\theta},
\end{equation}
that increases during the phase transition because $\theta$ is an angle that goes from $\theta_0=\frac{\pi}{2}$, to $\theta \rightarrow 0$.

The nonzero components of the Einstein tensor, are
\begin{equation}
G_{00} = - \frac{3}{\theta^2} , \qquad G_{ij} =  \frac{3}{\theta^2 } \,\delta{ij},
\end{equation}
with energy density $\rho$ and the pressure $P$
\begin{equation}
\rho(\theta) = \frac{1}{\pi G} \frac{3}{(\pi a_0)^2}, \qquad\qquad
P(\theta) = - \frac{1}{\pi G} \frac{3}{(\pi a_0)^2},
\end{equation}
such that the equation of state is
\begin{equation}
\frac{P}{\rho} =  - 1.
\end{equation}
Therefore, the universe suffered a vacuum expansion during preinflation with scale factor $a(t)=a_0\, e^{H_0 t}$, a  constant Hubble parameter $H_0=\frac{\dot{a}}{a}=\frac{2\sqrt{2}}{\pi\,a_0}$ and a potential $V(\varphi)= \frac{3}{8\pi G} H^2_0$, related to a scalar field $\varphi$. Since $\frac{\delta V}{\delta\varphi}=0$, therefore the background field background equation (\ref{dy}), result to be
\begin{equation}
\varphi{''}-\frac{2}{\theta} \varphi{'}=0,
\end{equation}
which has the solution
\begin{equation}
\varphi(\theta)= \varphi_0.
\end{equation}
This solution describes the background solution of the background field that drives a phase transition of the global geometry
from a 4D Euclidean space to a 4D hyperbolic space time. We shall suppose that at the beginning the universe has a size given by the Planck length: $a_0=L_p$, and the
total energy in the universe is given by the Planck mass $M_p$ (we shall consider that $c=1$).

\section{Boundary conditions in the minimum principle action with a Ricci flow: connections, covariant derivatives}

We consider the Einstein-Hilbert (EH) action for an arbitrary matter lagrangian density ${\cal L}_{\varphi}$
\begin{equation}
{\cal I} = \int d^4 x \sqrt{-g} \left[ \frac{R}{2\kappa}+ {\cal L}_{\varphi} \right],
\end{equation}
after variation, this expression is given by
\begin{equation}\label{delta}
\delta {\cal I} = \int d^4 x \sqrt{-g} \left[ \delta g^{\alpha\beta} \left( G_{\alpha\beta} + \kappa T_{\alpha\beta}\right)
+ g^{\alpha\beta} \delta R_{\alpha\beta} \right],
\end{equation}
where $\kappa = 8 \pi G$, ${{T}}_{\alpha\beta}$ is the background stress tensor
\begin{equation}\label{bt}
{{T}}_{\alpha\beta} =   2 \frac{\delta {{\cal L}_{\varphi}}}{\delta g^{\alpha\beta}}  - g_{\alpha\beta} {{\cal L}_{\varphi}},\\
\end{equation}
and ${{\cal L}_{\varphi}}$ is the Lagrangian density (\ref{lan}), that describes the background physical dynamics. The last term in (\ref{delta}) is very important because takes into account boundary conditions. When that quantity is zero, we obtain the well known Einstein's equations without cosmological constant $\lambda$. In the general case, the classical Einstein equations, with boundary conditions included, results to be
\begin{equation}
\bar{G}_{\alpha\beta} = - \kappa\, {T}_{\alpha\beta}, \label{e1}
\end{equation}
with $\bar{G}_{\alpha\beta} = {G}_{\alpha\beta} - \lambda\, g_{\alpha\beta}$.  Here, ${G}_{\alpha\beta}=R_{\alpha\beta}-\frac{1}{2}\,R\,g_{\alpha\beta}$ is the background Einstein tensor, where $R_{\alpha\beta}$ is the background Ricci tensor, $R=g^{\alpha\beta} R_{\alpha\beta}$ is the background scalar curvature, $g_{\alpha\beta}$ is the symmetric background metric tensor, and $G$ is the gravitational constant. We shall require that
\begin{equation}\label{flu}
g^{\alpha\beta} \delta R_{\alpha\beta}-\delta\Theta =
\left[\delta W^{\alpha}\right]_{||\alpha} - \left(g^{\alpha\epsilon}\right)_{||\epsilon}  \,\delta\Gamma^{\beta}_{\alpha\beta} +
\left(g^{\alpha\beta}\right)_{||\epsilon}  \,\delta\Gamma^{\epsilon}_{\alpha\beta}=0,
\end{equation}
such that $\delta W^{\alpha}=\delta
\Gamma^{\alpha}_{\beta\nu} g^{\beta\nu}-\delta\Gamma^{\epsilon}_{\beta\epsilon}
g^{\beta\alpha}$ and $\delta \Theta$ is the flux that cross the 3D-closed hypersurface due to the boundary terms included when we variate the action.

Additionally, we shall use a recently introduced extended manifold\cite{usf3} to describe quantum geometric spinor fields
$\hat{\Psi}^{\alpha}$, where the connections are
\begin{equation}\label{ga}
\hat{\Gamma}^{\alpha}_{\beta\nu} = \left\{ \begin{array}{cc}  \alpha \, \\ \beta \, \nu  \end{array} \right\}+ \hat{\delta{\Gamma}}^{\alpha}_{\beta\nu},
\end{equation}
where we shall consider that the varied connections
\begin{equation}\label{uch}
\hat{\delta{\Gamma}}^{\alpha}_{\beta\nu}=\epsilon\,\hat{\Psi}^{\alpha}\,g_{\beta\nu},
\end{equation}
describe the quantum displacement of the extended manifold with respect to the classical Riemannian background, which is described by the Levi-Civita symbols in (\ref{ga}).
In this work we shall consider the particular case where $\hat{\Psi}^{\alpha}$ are massive charged bosons of spin $1$, but in general, the operators $\hat{\Psi}^{\alpha}$ are the quantum spinor field components that represent rather bosons or fermions. Furthermore, $\epsilon$ is a constant related to self-interactions of the spinor field. The covariant derivative of the metric tensor and arbitrary vectors on the extended manifold, with self-interactions included, are respectively given by
\begin{eqnarray}
\hat{g}_{\beta\alpha\|\nu}& =&\nabla_{\nu} g_{\beta\alpha} -\epsilon\, \left(g_{\beta\nu} \hat{\Psi}_{\alpha} + \hat{\Psi}_{\beta} g_{\alpha\nu} \right) + 2\left(1-\xi^2\right) \hat{\Psi}_{\nu} \, g_{\alpha\beta}, \label{gg} \\
\left[\Upsilon^{\alpha}\right]_{||\beta}&=& \nabla_{\beta}\Upsilon^{\alpha} + \delta\Gamma^{\alpha}_{\epsilon\beta}\Upsilon^{\epsilon} - (1-\xi^2) \Upsilon^{\alpha}\hat\Psi_{\beta}
= \nabla_{\beta}\Upsilon^{\alpha} + \left(\epsilon \,\hat{\Psi}^{\alpha}\Upsilon_{\beta}-\Upsilon^{\alpha}\hat{\Psi}_{\beta}\right) + \xi^2 \Upsilon^{\alpha}\hat\Psi_{\beta},
\end{eqnarray}
where $\nabla_{\nu} g_{\beta\alpha}=0$ is the covariant derivative on the Riemann manifold, and $\|$ denotes the covariant derivative on the extended manifold.

If we require that $\epsilon=1/3$ in (\ref{uch}) and that the extended manifold takes the form of a Ricci flow: $\hat{\delta R}_{\alpha\beta}=\lambda\,\hat{\delta g}_{\alpha\beta}$. In this case we obtain in (\ref{flu}), that
\begin{equation}\label{gauge}
{g}^{\alpha \beta} \hat{\delta R}_{\alpha \beta}-\hat{\delta \Theta} =
\left[\hat{\delta W}^{\alpha}\right]_{||\alpha} - \left(g^{\alpha\beta}\right)_{||\beta}  \,\hat{\delta\Gamma}^{\epsilon}_{\alpha\epsilon} +
\left(g^{\epsilon\nu}\right)_{||\alpha}  \,\hat{\delta\Gamma}^{\alpha}_{\epsilon\nu}= \nabla_{\alpha}\,\hat{\delta W}^{\alpha}=-\nabla_{\alpha}\,\hat{\Psi}^{\alpha}=0,
\end{equation}
where we define the varied Ricci tensor using the extended Palatini identity\cite{pal},
\begin{equation}
\hat{\delta{R}}^{\alpha}_{\beta\nu\alpha}=\hat{\delta{R}}_{\beta\nu}= \left(\hat{\delta\Gamma}^{\alpha}_{\beta\alpha} \right)_{\| \nu} - \left(\hat{\delta\Gamma}^{\alpha}_{\beta\nu} \right)_{\| \alpha}.
\end{equation}
This implies that the flux $\hat{\delta\Theta}$ is given by
\begin{equation}
\hat{\delta \Theta} = \lambda\,{g}^{\alpha\beta}\,\hat{\delta g}_{\alpha\beta},
\end{equation}
where the cosmological constant, for the case of preinflation described in the earlier section, is $\lambda=3\,H^2_0$. The origin of $\lambda$ can be obtained from USF and is originated in fermion fields\cite{usf3}.

\subsection{Line elements and invariants of the theory}

To describe a non-commutative spacetime, we shall consider unit vectors are $4\times 4$-matrices: $\bar{\gamma}^{\alpha}$, that generate a globally hyperbolic spacetime. These matrices generate the background metric and we include the spinor information in the spacetime structure that can describe quantum effects in a relativistic framework. Once the basic elements are introduced, we can define the line elements of our theory:
\begin{equation}\label{line}
dx^2 \delta_{BB'}  = \left<B\right| \underrightarrow{\delta{X}} \overrightarrow{\delta{X}} \left| B'\right> , \qquad  d\phi^2 \delta_{BB'}=
\left<B\right| \underleftrightarrow{\delta\Phi} \overleftrightarrow{\delta\Phi}\left| B'\right>.
\end{equation}
The first expression in (\ref{line}) is a standard inner product, but the second one is a bi-vectorial product given by the expectation value of
\begin{equation}
\underleftrightarrow{\delta\Phi} \overleftrightarrow{\delta\Phi}= \frac{1}{4}\, \left(\hat{\delta\Phi}_{\alpha}\,\hat{\delta\Phi}_{\beta}\right)\,\left(\bar{\gamma}^{\alpha}\,\bar{\gamma}^{\beta}\right),
\end{equation}
where
\begin{equation}
\hat{\delta\Phi}_{\alpha}\,\hat{\delta\Phi}_{\beta} = \frac{1}{2} \left\{\hat{\delta\Phi}_{\alpha}, \hat{\delta\Phi}_{\beta}\right\} + \frac{1}{2} \left[\hat{\delta\Phi}_{\alpha}, \hat{\delta\Phi}_{\beta}\right], \qquad
\bar{\gamma}^{\mu} \bar{\gamma}^{\nu} = \frac{1}{2} \left\{\bar{\gamma}^{\mu}, \bar{\gamma}^{\nu}\right\} + \frac{1}{2} \left[\bar{\gamma}^{\mu}, \bar{\gamma}^{\nu}\right].
\end{equation}
The matrices that generates the spacetime $g_{\mu\nu} \,\mathbb{I}_{4\times 4} = (1/2) \left\{\bar{\gamma}_{\mu}, \bar{\gamma}_{\nu}\right\}$, are:
\begin{equation}
\bar{\gamma}^{\mu} = E^{\mu}_{\hskip.1cm\nu} \,\gamma^{\nu},
\end{equation}
where $E^{\mu}_{\hskip .1cm\nu}$ are the transformation matrices and the components $\gamma^{\mu}$ are given in the Weyl representation. These matrices\footnote{The Weyl representation of matrices in cartesian coordinates, are
\begin{eqnarray}
&& \gamma^0= \,\left(\begin{array}{ll}  0 & \mathbb{I} \\
\mathbb{I}  &  0 \ \end{array} \right),\qquad
\gamma^1=  \left(\begin{array}{ll} 0 &  -\sigma^1 \\
\sigma^1 & 0  \end{array} \right),  \nonumber \\
&& \gamma^2= \left(\begin{array}{ll} 0 &  -\sigma^2 \\
\sigma^2 & 0  \end{array} \right),  \qquad \gamma^3= \left(\begin{array}{ll} 0 &  -\sigma^3 \\
\sigma^3 & 0  \end{array} \right),\label{gamm}
\end{eqnarray}
where $\sigma^i$ are the Pauli matrices.}, describes a globally hyperbolic 4D-spacetime. Each component of spin $\hat{S}_{\mu}= s\,\bar{\gamma}_{\mu}$, is defined as the canonical momentum associated to the inner coordinate $\hat{\Phi}^{\mu}$. Therefore, the universal bi-vectorial invariant can be defined:
\begin{equation}\label{invariant}
\left<B\left| \underleftrightarrow{S} \overleftrightarrow{\Phi}\right|B\right> = \frac{1}{4} \left<B\left|\left( \hat{S}_{\mu} \hat{\Phi}_{\nu} \right) \left( \bar{\gamma}^{\mu} \bar{\gamma}^{\nu}\right)\right|B\right> =s \phi \,
\mathbb{I}_{4\times 4} = (2\pi n \hbar) \,\mathbb{I}_{4\times 4},
\end{equation}
with $n$-integer.

In the case of preinflation, which we are studying in this work the line element associated to the inner space is given by
\begin{equation}
d\phi^2 = \left( \bar{\gamma}_{\mu} \bar{\gamma}_{\nu}\right)\, d\phi^{\mu}\,d\phi^{\nu},
\end{equation}
and the transformation matrix is the diagonal one
\begin{equation}
E^{\mu}_{\hskip.1cm\nu} =  \left(\frac{\pi a_0}{2}\right) \,\frac{1}{\theta} \,\mathbb{I}_{4\times 4}.
\end{equation}
We define the variation of the metric tensor on the extended manifold, as
\begin{equation}
\hat{g}_{\beta\alpha\|\gamma} \,\hat{\delta X}^{\gamma} \left| B\right> = \delta g_{\beta\alpha} \left| B\right>,
\end{equation}
where such variation is defined with respect to the background Riemannian defined by the second kind Christoffel symbols. The fact that $\delta g_{\beta\alpha}\neq 0$ on the extended manifold, means that the geometry described on this manifold is elastic, rather than on the Riemann manifold in which there is null nonmetricity $\Delta g_{\alpha\beta}=0$. Now we are in conditions of defining the spinor fields, as the variation of the flux $\hat{\delta\Theta}$ with respect to the
inner coordinates: $\hat\Psi_{\alpha}=\frac{\hat{\delta\Theta}}{\hat{\delta{\Phi}}^{\alpha}}$
\begin{eqnarray}
 \hat{\Psi}_{\beta}\left(x^{\beta}|\Phi^{\alpha}\right)&=& \frac{i}{\hbar (2\pi)^4} \int d^4k \int d^4s \frac{\delta \left(\underleftrightarrow{S} \overleftrightarrow{\Phi}\right)}{\hat{\delta\Phi}^{\alpha}} \left[ A_{s,k}\, \hat{\Theta}_{k,s}(x^{\beta}) e^{\frac{i}{\hbar} \underleftrightarrow{S} \overleftrightarrow{\Phi}}
- B^{\dagger}_{k,s} \, \hat{\Theta}^*_{k,s}(x^{\beta}) e^{-\frac{i}{\hbar} \underleftrightarrow{S} \overleftrightarrow{\Phi}}\right], \nonumber
\end{eqnarray}
where
\begin{equation}
\frac{\delta }{\hat{\delta\Phi}^{\alpha}}\left(\underleftrightarrow{S} \overleftrightarrow{\Phi}\right) =  \left(2 g_{\alpha\beta}  \mathbb{I}_{4\times 4} - \bar{\gamma}_{\alpha} \bar{\gamma}_{\beta} \right) \hat{S}^{\beta} = 2 \hat{S}_{\alpha} - \bar{\gamma}_{\alpha} \,s =\,\bar{\gamma}_{\alpha} \,s ,
\end{equation}
where $s\,\mathbb{I}_{4\times 4}= \frac{1}{4} \hat{S}_{\beta} \bar{\gamma}^{\beta}$. Because we are considering massive charged bosons with  spin $s=\hbar$, due to the invariant (\ref{invariant}), the operators $e^{\pm\frac{i}{\hbar} \underleftrightarrow{S} \overleftrightarrow{\Phi}}$ applied on the background state $\left|B\right>$, will be invariant under $\phi= 2n\pi$-rotations
\begin{equation}
e^{\pm\frac{i}{\hbar} \underleftrightarrow{S} \overleftrightarrow{\Phi}}\left|B\right> \,= \,e^{\pm \,{i} \,\phi}  \,        \left|B\right>,
\end{equation}
where we have adopted the Heisenberg representation for these states. In this representation, the operators are evolving and states are squeezed.

\section{MM during preinflation}

We consider the more general quantum action to describe all possible symmetries in nature\cite{mb}:
\begin{equation}\label{act1}
I_{M} = \frac{c^4\,L_p\,t_p}{32\,\pi G} \, \int\, d^4x\,\int\, d^4\phi \,\sqrt{-g}\, \left<B\left| \left(\bar{\gamma}^{\mu} \bar{\gamma}^{\nu}\right)\, \left( \hat{\delta R}^{\alpha}_{\hskip.1cm\mu\nu\alpha} +
\hat{\delta R}^{\alpha}_{\hskip.1cm\alpha\mu\nu}\right) \right|B\right>,
\end{equation}
where
\begin{equation}
\hat{\delta{R}}^{\alpha}_{\hskip .1cm\alpha\beta\nu}= \left(\hat{\delta\Gamma}^{\alpha}_{\hskip .1cm\alpha\nu} \right)_{\| \beta} - \left(\hat{\delta\Gamma}^{\alpha}_{\hskip .1cm\alpha\beta} \right)_{\| \nu}.
\end{equation}
Here, $c$ is the light velocity, $t_p$ is the Planck time and $L_p$ is the Planck length. The matrices $\bar\gamma^{\mu}$, comply with the Clifford algebra:
\begin{equation}
\bar{\gamma}^{\mu} = \frac{\bf{I}}{3!}\,\epsilon^{\mu}_{\,\,\alpha\beta\nu} \bar{\gamma}^{\alpha}\bar{\gamma}^{\beta}  \bar{\gamma}^{\nu} , \qquad \left\{\bar{\gamma}^{\mu}, \bar{\gamma}^{\nu}\right\} =
2 g^{\mu\nu} \,\mathbb{I}_{4\times 4}, \nonumber
\end{equation}
where ${\bf{I}}={\gamma}^{0}{\gamma}^{1}{\gamma}^{2}{\gamma}^{3}$ is the pseudoscalar, $\mathbb{I}_{4\times 4}$ is the identity matrix, and we define $\noindent{{\epsilon}^{\mu}_{\alpha\beta\nu}=g^{\mu\rho}\epsilon_{\rho\alpha\beta\nu}}$, such that $\epsilon_{\rho\alpha\beta\nu}$ are the Levi-Civita symbols. In order to describe the quantum action due to MM, we must consider only massive and charged antisymmetric contributions of (\ref{act1}), which could remain stable. This means that they would not radiate as electromagnetic fields and therefore only we must consider the $\hat{\cal{M}}_{\mu\nu}$-contributions
\begin{equation}\label{act3}
I_{M} = -\frac{c^4\,L_p\,t_p}{32\,\pi G}\,  \int\, d^4x\,\int\, d^4\phi \,\sqrt{-g}\, \left<B\left|  \frac{3}{2} \left[\bar{\gamma}^{\mu}, \bar{\gamma}^{\nu}\right] \hat{\cal{M}}_{\mu\nu} \right|B\right>,
\end{equation}
where $\hat{\cal{M}}_{\beta\nu}$ are the components of a purely antisymmetric operator given by
\begin{equation}
\hat{\cal{M}}_{\beta\nu}  =  \alpha_1\,  \left[ \hat{\Psi}_{\beta}, \hat{\Psi}_{\nu}\right], \label{n2}
\end{equation}
where $\alpha_1= \frac{(5-6\xi^2)}{36}$. We must remember that we are considering $\epsilon=1/3$ in (\ref{uch}). To quantise the massive and charged spin $1$-bosons, we must take into account that they are geometric quantum spinor fields, and could be responsible for the MM at the beginning of the universe. If we consider that this phenomena became from a condensate of very massive and charged boson fields, described by the strength field tensor $\hat{\cal{M}}^{\beta\nu}$, which is conserved on the extended manifold, then this field must fulfil the expression
\begin{equation}\label{cons}
\left(\hat{\cal{M}}^{\beta\nu}\right)_{\|\nu}=0,
\end{equation}
that can be rewritten in terms of Riemannian covariant derivatives for $\epsilon=1/3$:
\begin{equation}\label{mm}
\nabla_{\beta} \hat{\cal{M}}^{\beta\nu}+ (1-\xi^2)\,\left[\hat{\Psi}_{\beta},\hat{\cal{M}}^{\beta\nu}\right]+\frac{1}{3}\,\hat{\Psi}_{\beta}\,\hat{\cal{M}}^{\beta\nu}=0.
\end{equation}
Here, we can identify the magnetic current components $J^{\nu}_{(m)}$
\begin{equation}
J^{\nu}_{(m)}=-\frac{1}{\mu_0}\,\left[(1-\xi^2)\,
\left[\hat{\Psi}_{\beta},\hat{\cal{M}}^{\beta\nu}\right]+\frac{1}{3}\,\hat{\Psi}_{\beta}\,\hat{\cal{M}}^{\beta\nu}\right],
\end{equation}
in order to obtain $ \nabla_{\beta} \hat{\cal{M}}^{\beta\nu} =\mu_0\,J^{\nu}_{(m)}$. In particular, the $0$-component give us
the magnetic density of charge: $J^{0}_{(m)}=c\,\rho_{m}$. The equation (\ref{mm}), after
taking into account the last expression in (\ref{gauge}), takes the form
\begin{equation}
\left(\nabla_{\nu} \hat{\Psi}^{\mu}\right) \,\hat{\Psi}^{\nu} +\alpha_2\,\hat{\Psi}_{\mu} \left[\hat{\Psi}^{\mu},\hat{\Psi}^{\nu}\right]
+\alpha_3\,\left[\hat{\Psi}^{\mu},\hat{\Psi}^{\nu}\right]\,\hat{\Psi}_{\mu}=0,
\end{equation}
with
\begin{equation}
\alpha_2=\frac{(4-3\xi^2)}{4}, \qquad \alpha_3=(1-\xi^2).
\end{equation}
We remember that the $\nabla_{\alpha}$ denotes the covariant derivative on the background Riemann manifold, and therefore describes derivatives with respect to the Levi-Civita connections \begin{tiny}$\left\{ \begin{array}{cc}  \alpha \, \\ \beta \, \nu  \end{array} \right\}$\end{tiny}. The quantisation expression
for the case of massive and charged spin $1$-bosons, which are the interest in this work, are
\begin{equation}\label{con}
\left< B\left| \left[\hat{\Psi}_{\mu}({\bf x}, {\bf \phi}), \hat{\Psi}_{\nu}({\bf x}', {\bf \phi}') \right]\right|B \right>
= \frac{s^2}{2 \hbar^2 } \left[\bar\gamma_{\mu} , \bar\gamma_{\nu}\right] \,  \sqrt{\frac{\eta}{g}} \,\,\delta^{(4)} \left({\bf x} - {\bf x}'\right) \,\delta^{(4)} \left({\bf \phi} - {\bf \phi}'\right),
\end{equation}
where $s=\hbar$, $\sqrt{\frac{\eta}{g}}$ is the squared root of the ratio between the determinant of the Minkowsky metric: $\eta_{\mu\nu}$, and the metric that describes the background: $g_{\mu\nu}$. This ratio describes the inverse of the relative volume of the background manifold. Using
(\ref{con}) in (\ref{act3}), we obtain that the action related to MM in the preinflationary universe, is
\begin{equation}
{\cal I}_M = \frac{\left(5-6\xi^2\right)}{16}\,\hbar.
\end{equation}
Therefore, we can interpret that the total mass due to magnetic monopoles $m_M$, in the initial state of the universe with a Planck size, is:
\begin{equation}
m_M = \frac{\left(5-6\xi^2\right)}{16}\,M_p \, \geq 0.
\end{equation}
This means that the coupling constant $\xi^2$ will have the cut
\begin{equation}
\xi^2 < \,\frac{5}{6} ,
\end{equation}
so that $\alpha_2>0$ and $\alpha_3>0$. If we consider that $\xi$ is related to the ratio between the magnetic monopole charge and the Planck charge: $\xi^2=(q_M/q_p)^2$, then we obtain the following condition
\begin{equation}
\left(\frac{q_M}{q_p}\right)^2 = \left(\frac{5}{6} - \frac{8 \,m_M}{3 \,M_p}\right) \,<\,\frac{5}{6},
\end{equation}
where $q_p=e/\sqrt{\alpha}$, $e$ is the electron charge and $\alpha=1/137.03599$ is the fine structure constant. Therefore, finally we obtain that the
fine structure constant is given by
\begin{equation}
\alpha= \frac{5}{6}  \left(1- \frac{16 \,m_M}{5 \,M_p}\right) \,\left(\frac{e}{q_M}\right)^2, \label{mg}
\end{equation}
and therefore $m_M < (5/16)\,M_p$, in order to $q_M$ to be real, and the cut for the total magnetic charge due to MM is $q_M < 10.6848\,e$.

\subsection{Universe with MM with proton mass $m_P$ and charge $q_M=\pm e$, at the Planck era}

Notice that, for MM with total magnetic charge $q_M$, and a total magnetic mass $m_M$, there is the cut  $m_M < 5\,M_p/16$. Due to the fact these monopoles have a topological origin and are considered quantum in nature, hence they could have a small mass. For example, in the case of quantum monopoles with a proton mass: $m_P=7.68\times 10^{-20}\,M_p$, there would be $n=4.03\times 10^{18}$ completing a total mass $m_M=n\,m_P=0.3097\,M_p$ of quantum MM at the Planck era. Because they are topologically stable configurations of massive and charged spin $1$-bosons fields, MM with a proton mass could be considered as very good candidates to explain the origin of DM in the early universe at Planck
scales with a $ 30.97\,\%$ of the total mass in the universe, which is considered $M_p$.

\subsection{Universe with lights MM of fractional magnetic charge, at the Planck era}

\begin{figure}
  \centering
    \includegraphics[scale=0.8]{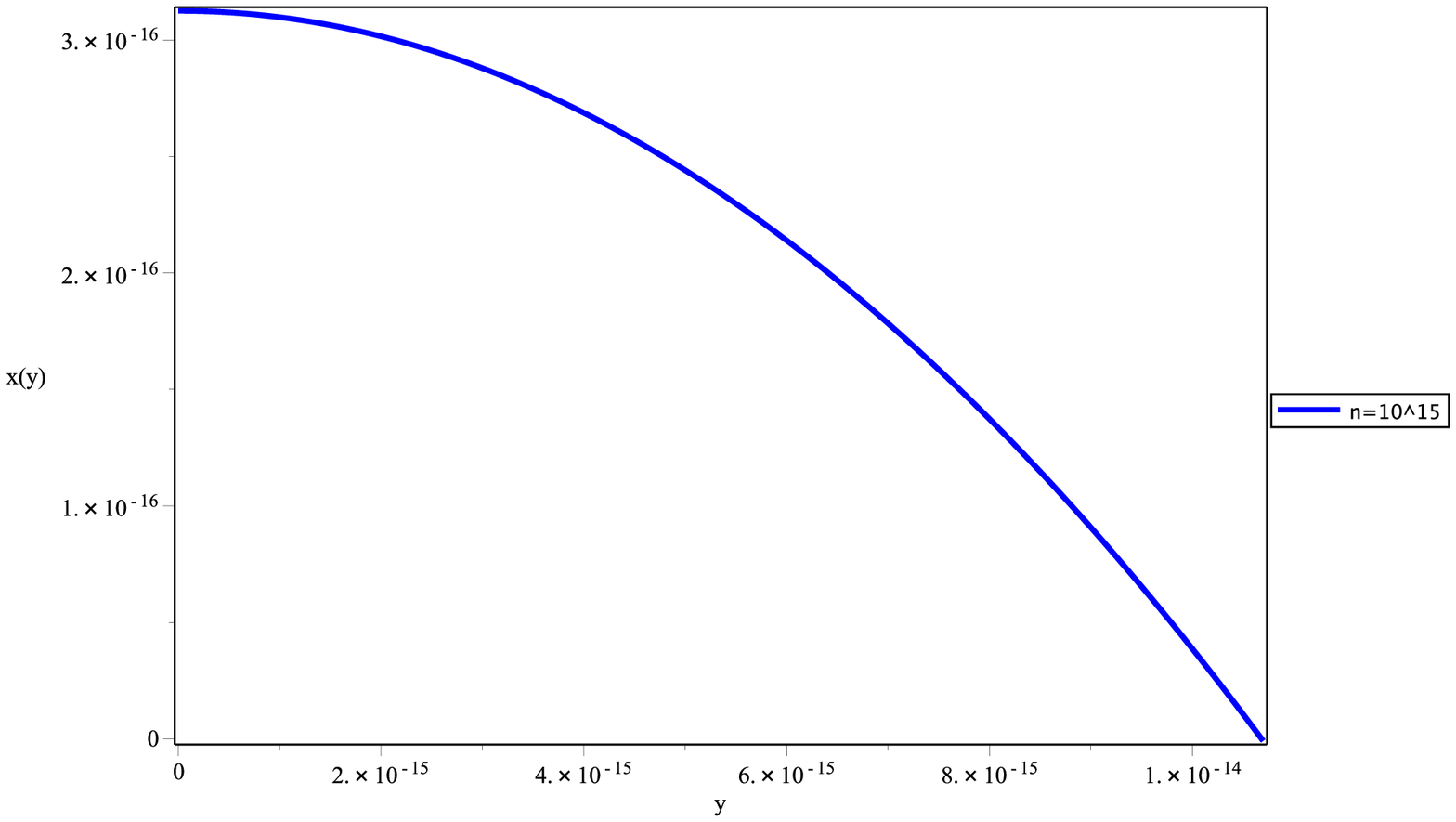}
  \caption{Plot of $x(y)$ with $n=10^{15}$ magnetic monopoles (MM), for $x<10^{-14}$ and $n=10^{15}$. Notice that $0< \,x < 3\times\,10^{-16}$, which means that MM have masses less than $3.6\times 10^3$ GeV.}
  \label{f1gi120}
\end{figure}

We can consider another interesting case which can be experimentally studied in a collider. This is the case of
an initial universe with $n$ little MM that have a total mass $m_M=n\,x\,M_p$ and total charge $q_M=\pm n\,y\,e$, for $x<1$ and $y<1$.  They are called milli-magnetically charged particles in \cite{mili} and are of direct interest to searches performed in the ATLAS\cite{B} and MoEDAL\cite{A} experiments at the Large Hadron Collider (LHC) at CERN. In this case, from the Eq. (\ref{mg}), we obtain
\begin{equation}\label{ei}
0\,<x \equiv \frac{685- 6\, n^2 \,y^2}{2192\,n} \,<1.
\end{equation}
Here, the mass of each MM is $x\,M_p < M_p$ and its magnetic charge will be $y\,e < e$. From the first inequality in (\ref{ei}), we obtain that the
number of MM at the Planck scale is
\begin{equation}
n = \sqrt{\frac{685}{6\,y^2}},
\end{equation}
which is expected to be $n \gg 1$ to MM with fractional magnetic charge: $y < 1$. In the Fig. (\ref{f1gi120}) we have
plotted $x(y)$ with $n=10^{15}$ MM, for $y<10^{-14}$ and $n=10^{15}$. For this $n$-value we obtain particles with masses less than $3.6\times 10^3$ GeV (because $M_p=1.2\times 10^{19}$ GeV).

\section{Final comments}

We have studied the origin of primordial MM during the primordial universe described through a preinflationary model. These monopoles are originated from quantum vacuum excitations on the background Riemann manifold. From the mathematical point of view, the quantum vacuum excitations described by (\ref{act3}), take into account massive and charged spin $1$-bosons fields that came exclusively from the structure of quantum space-time. The tensor field $\hat{\cal{M}}_{\mu\nu}$ is purely antisymmetric but this antisymmetry is only due to this space-time structure, rather than torsion. There is no torsion in the USF theory because $\hat{\delta{\Gamma}}^{\alpha}_{\beta\nu}=\hat{\delta{\Gamma}}^{\alpha}_{\nu\beta}$. A very important result is the expression (\ref{mg}), that relates the fine-structure constant with the magnetic charges and mass, for an universe created with Planck energy $M_p$ and a total Planck charge $q_p$. To conclude, quantum MM could explain the origin of DM in the universe, because they are created by massive and charged spin $1$-bosons which are topologically stable. Topological stability is provided by the structure of the quantum space-time which is preserved with the expansion of the universe.

In particular, when we study the universe at the Planck era, it is obtained the quantisation rule (\ref{mg}) from the quantisation of the action (\ref{act3}). This rule provides the possible charges and masses of MM. From this expression
it is obtained that the maximum MM-magnetic charge is $q_M < 10.6848\,e$. Therefore, MM with bigger magnetic charges are excluded by the USF theory, and is avoided the problem in the calculation of $68.5\,e$-MM cross section production, that appears with the Dirac rule for quantisation of charges\cite{mpi}. To conclude, we have analysed a case that can be of experimental interest in the ATLAS\cite{B} and MoEDAL\cite{A} experiments; we demonstrate that very small MM with mass of up to $3.6\times 10^3$ GeV, can exist with charges less than $10^{-14}\,e$.

\section*{Acknowledgements}

\noindent
\noindent M. B. acknowledges CONICET, Argentina (PIP 11220150100072CO) and UNMdP (EXA955/20) for financial support.

\newpage

\end{document}